\documentclass{jpp}

\usepackage{graphicx}
\usepackage{natbib}
\usepackage[colorlinks,urlcolor=blue,citecolor=blue,linkcolor=blue]{hyperref}
\usepackage{amsmath}

\usepackage{amssymb}
\usepackage{mathtools}
\usepackage{lscape}
\usepackage{color}  
\usepackage{rotating}
\usepackage{booktabs}
\usepackage{caption}
\usepackage{ulem}
\usepackage{psfrag}
\usepackage{pstool}

\ifCUPmtlplainloaded \else
  \checkfont{eurm10}
  \iffontfound
    \IfFileExists{upmath.sty}
      {\typeout{^^JFound AMS Euler Roman fonts on the system,
                   using the 'upmath' package.^^J}%
       \usepackage{upmath}}
      {\typeout{^^JFound AMS Euler Roman fonts on the system, but you
                   dont seem to have the}%
       \typeout{'upmath' package installed. JPP.cls can take advantage
                 of these fonts, if you use 'upmath' package.^^J}%
      }
  \else
  \fi
\fi

\ifCUPmtlplainloaded \else
  \checkfont{msam10}
  \iffontfound
    \IfFileExists{amssymb.sty}
      {\typeout{^^JFound AMS Symbol fonts on the system, using the
                'amssymb' package.^^J}%
       \usepackage{amssymb}%
         \let\leq=\leqslant
         \let\geq=\geqslant
      }{}
  \fi
\fi

\ifCUPmtlplainloaded \else
  \IfFileExists{amsbsy.sty}
    {\typeout{^^JFound the 'amsbsy' package on the system, using it.^^J}%
     \usepackage{amsbsy}}
    {\providecommand\boldsymbol[1]{\mbox{\boldmath $##1$}}}
\fi

\newcommand{\be}{\begin{equation}} 
\newcommand{\ee}{\end{equation}}
\newcommand{\nn}{\mbox{} \nonumber \\ \mbox{} }
\newcommand{\ba}{\begin{eqnarray}}
\newcommand{\ea}{\end{eqnarray}}
\newcommand{\om}{\omega}

\newcommand{\Bf}{{magnetic field}}
\newcommand{\Bfs}{{magnetic fields}}
\newcommand{\Ef}{{electric  field}}

\newcommand{\NS}{neutron star}
\newcommand{\NSs}{{neutron stars}}
\newcommand{\EM}{electromagnetic}

\newcommand{\ms}{magnetosphere}
\newcommand{\mss}{magnetospheres}

\newcommand{\LC}{light cylinder}
\newcommand{\Lf}{Lorentz factor}
\newcommand{\Lfs}{Lorentz factors}
\newcommand{\curl}{{\rm curl\, }}

\begin{document}

\title{Coherence constraints on physical parameters at bright radio sources and FRB emission mechanism}

\author{Maxim Lyutikov,$^1$ Mohammad Z. Rafat,$^2$ }
\affiliation{$^1$ Department of Physics and Astronomy, Purdue University, 
 525 Northwestern Avenue,
West Lafayette, IN
47907-2036, USA; lyutikov@purdue.edu\\
$^2$SIfA, School of Physics, The University of Sydney, NSW 2006, Australia;
mohammad.rafat@sydney.edu.au
}

 \maketitle

\begin{abstract}
We discuss physical constrains that observations of high  brightness temperature coherent  radio emission, with brightness temperatures as high as  $T_b \sim 10^{35}$ K,  impose on the plasma parameters at  relativistically moving  astrophysical sources. High  brightness temperatures imply a minimal plasma energy density at the source. Additional important constraints come from the fact that resonantly emitting particles lose most of their  energy to non-resonant inverse Compton and synchrotron processes. 
Overall, we find that coherence constraints can be accommodated by plasma parameters in the \mss\ of \NSs.

We also interpret recent observations of high-to-low frequency  drifting features in the spectra of repeating FRBs as analogues of type-III Solar radio bursts produced by reconnection plasma beams within \mss\ of highly magnetized \NSs.
\end{abstract}

\section{Introduction} 

A number of astrophysical sources show high brightness temperatures, reaching, in case of pulsars and fast radio bursts (FRBs), values of $\sim 10^{35}$ K \citep[\eg][]{1977puls.book.....M,Melrose00Review,2007Sci...318..777L}, and as high as  $\sim 10^{40}$ K in extreme cases \citep[\eg][]{2004ApJ...616..439S}. These observations imply physical constraints at the emitting plasma that we discuss  below.

Though high brightness bursts from Crab and other pulsars have been known for a long time \citep{1968Sci...162.1481S}, recent  observations of  mysterious FRBs \citep{2007Sci...318..777L,2019A&ARv..27....4P,2019arXiv190605878C}  impose new constraints  on the physical properties at the source.
One of the more constraining limitations come from recent identification of a repeating FRB \citep{2016Natur.531..202S}. The implied luminosity of $\sim 10^{40}$ erg s$^{-1}$, taken with short duration $\sim $ few milliseconds  and large  distances $\sim 1 $ Gpc  implies  very high energy density at the source. 
 \cite{2017ApJ...838L..13L} (see also  \cite{2016MNRAS.462..941L})   argued that these constraints 
limits
 the {\it loci} of the radio emission generation to {\NS}s' \mss. 
 
 In this paper we discuss  more thoroughly what limitation on the   physical parameters at the relativistically moving source (\eg  density and \Bf) can be obtained from the  
the  observations of high brightness radio emission.  In \S \ref{General} we discuss general relations, without limiting ourselves to any particular physical location, while in \S \ref{NS} we apply these to the  {\NS}s. Finally, in \S \ref{drifts} we discuss implications of recent observations of frequency drifts in FRBs.

\section{Limits on physical parameters at the sources of high brightness radio emission} 
\label{General} 

\subsection{Radiation energy   density}
\label{General1} 
The extremely high brightness temperatures in astrophysical radio sources, most likely, involve relativistic plasma so that in the center of momentum frame the spread of the \Lfs\ $\gamma$ is large, $\gamma \gg 1$. The plasma is likely to be in a \Bf\ and additionally the center of momentum frame may move with respect to the observer with relativistic bulk streaming \Lf\ $\gamma_s \gg 1$. Let us first neglect the effects of the bulk motion, to be added later in \S \ref{realt}.

Consider a radiation source with observed spectral flux density $ S_\nu $ at distance $ r $ from the observer. For an isotropic source we have, Appendix~\ref{app:sec:isotropic},
\begin{equation}\label{eq:Snu0}
    S_\nu = \pi \frac{R^2}{d^2}I_\nu,
\end{equation}
where $ R $ is the source radius and $ I_\nu $ is its spectral brightness. For an unresolved source we may estimate its radius from its variability time $ \tau $ as $ R \approx c\tau $, Appendix~\ref{app:sec:unresolved}, so that $ S_\nu = \pi (c\tau)^2 I_\nu/d^2 $. For an anisotropic source~\eqref{eq:Snu0} is further modified, Appedix~\ref{App:sec:anisotropic}, to $ S_\nu = \pi\theta_b^2 (c\tau)^2 I_\nu/d^2 $ where $ \theta_b $ is half-cone angle of emission region. We may then write, using~\eqref{eq:Snu0} and~\eqref{eq:final},
\begin{equation}
    u_r = \frac{\Delta\nu S_\nu d^2}{c^3\tau^2} = \frac{S d^2}{c^3\tau^2},
\end{equation}
where $ u_r $ is the energy density, $ S $ is the total flux and $ \Delta\nu $ is the frequency range corresponding to central frequency $ \nu $ with $ \Delta/\nu/\nu \ll 1 $. The brightness temperature for unresolved unpolarized source is defined as 
\be
k_B T_b = \frac{c^2 I_\nu}{2 \nu^2}.
\ee
We may then relate the  energy density of radiation at the source  $u_r$ to the observed  brightness temperature, using~(\ref{eq:final}), as
\be
    u_r = \left(  \frac{\Delta \nu}{\nu}  \theta_{\rm b}^2  \right)  \frac{2\pi  k_B T_b}{\lambda^3} 
    \label{ur1}
\ee
where $\lambda= \nu/c$ is the wavelength, ${\Delta \nu}/{\nu}$ is the fractional bandwidth of the receiver, and $\theta_{\rm b}$ is a typical beaming angle of the radiation at the source.    The factor in parenthesis is smaller than unity, hence it {\it relaxes}  constraints on the plasma parameters. This factor can in fact be much smaller than unity, with $\theta_{\rm b}^2  \sim 1/\gamma^2$.

To avoid complications about the unknown microphysics of the coherent emission, below we incorporate the factor in parenthesis into brightness temperature 
\be
    \overline{T}_b \equiv    \left(  \frac{\Delta \nu}{\nu}  \theta_{\rm b}^2  \right)   T_b.
\ee
The observed brightness temperature is $ T_{\rm b} $ while we use $ \overline{T}_{\rm b} $ to infer intrinsic plasma parameters. Relation (\ref{ur1}) then becomes
\be
    u_r =   \frac{2\pi  k_B    \overline{T}}{\lambda^3} 
    \label{ur2}
\ee
% Thus, for isotropically emitting source the brightness temperature is approximately the energy density of radiation within a volume of wavelength cubed. 
Relation (\ref{ur2}) is one of the key results of the paper: it connects the observed brightness temperature $T_B$, and the properties of the detector ${\Delta \nu}/{\nu} $ and $\lambda$ to the radiation energy density at the source (parametrized by beaming angle $\theta_b$.

\subsection{Required plasma density and \Bf}

Radiation energy   density at the source  $u_r$ should be a fraction $\eta_r \leq 1$ of emitting particles' energy density,
\be\label{eq:ur1b}
u_{r}=\eta_r \gamma n  m_e c^2,
\ee
where $\gamma$ is the random \Lf\ of particles at the source. Values of $\gamma$ and $n$ refer only to the coherently emitting particles, not the total energy/density. Hence equating~\eqref{ur1} and~\eqref{eq:ur1b} implies
% \be
% k_B T_b ={\eta_r}  \left(  \frac{\Delta \nu}{\nu}  \theta_{\rm b}^2  \right)  ^{-1}    \frac{ \gamma n m_e c^2 \lambda ^3    }{2 \pi} 
% \label{01}
% \ee

% Using $\eta_r \leq  1$, the condition
% (\ref{01}) becomes
\be 
     \gamma  n  
        = \frac{1}{\eta_r}  \frac{2\pi k_B \overline{T}_b}{m_e c^2 \lambda ^3}
        \geq   \frac{2\pi k_B \overline{T}_b}{m_e c^2 \lambda ^3},
 \label{03}
 \ee 
 where the inequality follows as $ \eta_r \leq 1 $. This is the requirement on the plasma energy density at the source given the observed brightness temperature.

Given the  radiation  flux and the  radiation energy density  (\ref{ur1}) one can calculate the equipartition \Bf\ $B_{eq} $ \citep{2017ApJ...838L..13L,2016MNRAS.462..941L}) by equating $ u_r $ with the magnetic energy density $ u_B = B_{eq}^2/8\pi $:
\be
    B_{eq} 
        % = \left(\frac{8\pi \Delta\nu F_\nu d^2}{c^3\tau^2}\right)^{1/2}
        % =\left(16 \pi^3 k_B \frac{\overline{T}_b}{\lambda^3}\right)^{1/2}
        = \left(\frac{16 \pi^3 k_B}{c^3} \nu^3 \overline{T}_b\right)^{1/2}
        \approx 5\times10^8  \left(\frac{\nu}{1~{\rm GHz}}\right)^{3/2}\left(\frac{\overline{T}_b}{10^{35}~{\rm K}}\right)^{1/2}\, {\rm G}
\label{Beq}
\ee
The equipartition \Bf\ is the lower estimate on the \Bf\ at the source, so that we can parametrize $B= b_{eq} B_{eq}$ with  $b_{eq} \geq 1$. 

The \Bf\ (\ref{Beq}) can occur, first of all,  in the \NS\ \mss, but also in the \mss\ of white dwarfs and stellar mass black holes. (See also \S \ref{a1} for another  requirements on the \Bfs\ at the source.)

\subsection{Relativistically moving  variable  sources}
\label{realt}

%THIS SECTION IS FROM LYUTIKOV, SECTION FROM Rafat IS BELOW, \S \ref{realt1}. I THINK LORENTZ TRANSFORMATIONS ARE CORRECT HERE.

If an unresolved 
source of emission is moving relativistically  with \Lf\ $\Gamma$ (and the corresponding Doppler factor with respect to the observer $\delta$),  and varies with a time scale $\tau'$ in the plasma rest frame, then using Lorentz transformations \citep[\eg][]{2013LNP...873.....G}
\ba &&
F_\nu  = \delta ^3 F_\nu^\prime
\nn &&
\nu F_\nu  = \delta ^4  (\nu F_\nu)^\prime
\nn &&
\nu \propto \delta \nu^\prime
\nn &&
\tau =\delta ^{-1} \tau' 
\nn &&
n= \Gamma n'
\nn &&
T_b= \delta^3 T_b'
\nn &&
u_{r}^\prime= \frac{(\nu F_\nu)^\prime d^2}{c^3 \tau^{\prime,2}}\delta ^2
\label{LF}
\ea
(primes denote quantities measured in the plasma rest frame),

 Hence
 \be
 {k_B \overline{T}_b} = 
 \eta_r \frac{ {\gamma \lambda ^3  m_e c^2 n' } }{2\pi} \delta^{4} 
 \label{05}
 \ee
 Or
 \be
 \gamma \delta^{4}   n'  > {2\pi} \frac{k_B \overline{T}_b}{ m_e c^2 \lambda^3}
 \label{cond}
 \ee
Expression (\ref{cond}) is a requirement on internal plasma parameters, $\gamma$ and $n'$, and bulk motion Doppler factor  $\delta$ in terms of the observed brightness temperature  $T_b$ and wavelength $\lambda$.

Inverting (\ref{cond}), we can put a limit on the highest possible frequency where coherent emission  of given brightness $T_b$ can be observed:
\be
\nu_c \leq \left(\frac{m_e n' c^5}{k_B \overline{T}_b} \right)^{1/3}  \gamma^{1/3} \delta ^{4/3}
\ee
Using definition of the brightness temperature this translates to
\be
\nu_c \leq \frac{n' m_e c^5 \tau^2 }{d^2 F_\nu}\gamma \delta^4 
\ee
We stress that $n'$ is the plasma (emitting leptons') density in the rest frame; in the observer frame $n= \Gamma n'$.

If the plasma is moving along \Bf, the estimate (\ref{Beq}) remains valid in the plasma rest frame. (Otherwise $B'= B/\Gamma$).  This immediately implies that $\sim 10^{35}$K brightens temperatures must come from  {\NS}s' \mss\  \citep{2016MNRAS.462..941L}.

Using parametrization of the \Bf\ at the source (\ref{Beq}) and the enthalpy density $\approx \gamma n' m_e c^2$,  the magnetization parameter $\sigma$  \citep{1984ApJ...283..694K} is
\be
\sigma = \frac{B^2}{4 \pi \gamma n' m_e c^2} \approx  \eta_r b_{eq}^2 \delta ^4 
\ee
In a relativistic plasma we expect $\sigma \geq 1$.  This requires (i) super-equipartition \Bf\  $b_{eq} \geq 1$; (ii) relativistic  Doppler motion $\delta \geq 1$; (iii)   not highly matter dominated regime -  $\eta_r $ not too small.

\section{Nonlinearity parameter $a$ and ``normal'' radiative losses}
\label{a1}

A particle emitting coherent radio emission will also experience  ``normal'' losses due to synchrotron and/or Inverse Compton processes. Under certain conditions these normal losses may dominate over  losses to coherent emission, further containing the properties of the emission region.

The conventional  \EM\ wave intensity parameter  \citep[\eg][]{LLII} evaluates to
\ba &&
a \equiv  \frac{ e E'}{2\pi m_e c \nu'}=\frac{e \sqrt{\nu k_B \overline{T}_b}}{m_e c^{5/2} \delta}=3 \times 10^5 \nu_{GHz}^{1/2} T_{b, 35}^{1/2} \delta^{-1}
\nn &&
\frac{E^{\prime, 2}}{4\pi}= u_r'
\label{a}
\ea
($E'$ is the \Ef\ of the wave in the rest frame). Conventionally, {\it in the absence of strong external \Bf}, parameter $a$ is a dimensionless  transverse momentum of a particle in the \EM\ wave, $a \sim p_\perp/(m_e c)$.

If a particle oscillates in the \EM\ fields of the coherent wave with amplitude $a$,
then in addition to the energy losses to the emission of coherent waves it will also suffer Inverse Compton  (IC) and  synchrotron losses (if \Bf\ is present). Typically  those ``normal'' losses will dominated over the ``coherent'' losses. For example,
if an external \Bf\ is present, the  particle  will also emit  synchrotron radiation. The  synchrotron radiation decay times
become shorter than pulse duration in the plasma frame for
\be
\overline{T}_b \geq \frac{1}{k_B} \frac{ m_e ^{8/3} c^7}{ e^{10/3} \nu^{7/3} \tau^{2/3}} = 
10^{29} b_{eq}^{-4/3}  \nu_{GHz}^{-7/3} \tau_{-3} ^{-2/3}\, {\rm K}
\label{Tmax}
\ee
 In case of IC losses, the estimate (\ref{Tmax}) remains valid with $b_{eq}$ set to unity.

Thus, under ``normal'' circumstances,  for brightness temperatures exceeding (\ref{Tmax}) the coherent \EM\ wave cannot be sustained: normal losses will drain particles' energy before it has time to emit a coherent wave.

There is an important caveat to the above statement. In large \Bfs, with $\om_B \geq 2\pi \nu'$ where $\om_B = e B/(m_e c)$ and $\nu'$ is the wave frequency in the source frame,  the nature of the particle's motion in the field  of the \EM\ wave changes: instead of   oscillations   in the \Ef\ of the wave with the dimensionless momentum $a$ a particle experiences slow drift with velocity $v_d/c \sim E'/B$ (here $E'$ is the \Ef\ of the wave, while $B$ is the external \Bf). This condition translates to
\be
B \geq \frac{ m_e c^2}{e \lambda \delta} \approx 100\nu_{GHz} \delta^{-1}\, {\rm G} 
\ee
Thus, the  presence of high \Bfs\ in high brightness relativistic sources is required to avoid large radiative losses of coherently emitting particles to IC  and synchrotron processes.

\section{Neutron stars' \mss}
\label{NS}

The relations derived above involve the source number density $n'$ and {\Lf}s $\gamma$ and $\gamma_s$. To proceed further we can parametrize the number density to the expected ones, particularly in the case of \NS\ \mss. Two parameterizations are possible: (i) pulsar-like, 
 normalizing the rest frame density to the GJ density \citep{GoldreichJulian}
\be
n' _{pulsar}= \kappa \frac{\Omega B}{2 \pi e c \gamma_s}
\label{n11}
\ee
where $\kappa\sim 10^3-10^6$ is the observer frame multiplicity \citep{1977ApJ...217..227F,2010MNRAS.408.2092T} and $\Omega$ is the \NS\ spin;
and (ii)  magnetar-like \citep{tlk02}
\be
n' _{magnetar}=  \Delta \phi \frac{B }{2\pi e r \gamma_s}
\label{n22}
\ee
(the last comes from equating $\curl B \sim \Delta \phi B/r$ to $(4\pi/c) 2 n e c$,  $\Delta \phi$ is a typical twist angle in the magnetar \ms.) 

For pulsar-like parametrization (\ref{n11}) the condition (\ref{cond}) gives
\ba &&
\eta_r = \frac{r_e k_B T}{\kappa  \gamma \lambda^3  \Omega \om_B m_e \gamma_s^3}
\nn &&
\kappa  \gamma \gamma_s^3 > \frac{r_e k_B T}{ \lambda^3  \Omega \om_B m_e }=
5\times 10^{18} \nu_{GHz}^3 P_0^{4} T_{b, 35} \left( \frac{B}{B_Q}\right)^{-1} \left( \frac{r}{R_{LC}}\right)^{3}
\ea
where $B_q= 4 \times 10^{13}$ Gauss is the quantum critical \Bf\ and 
we estimated $\delta \sim 2 \gamma_s$  and normalized to period $P_0=$ one second; $R_{LC}= c/\Omega$ is the light cylinder radius.
In the extreme case, a  millisecond pulsar with quantum \Bf\ producing radiation near the \LC, it is required that
\be
\kappa  \gamma \gamma_s^3 > 6 \times 10^6 T_{b, 35}
\ee
This is not too constraining, since we expect $\kappa \sim 10^4-10^6$ and $\gamma \sim \gamma_s \sim 10^3-10^4$ \citep[\eg][]{Hibschman,2010MNRAS.408.2092T}.

Similarly, for Crab giant pulses 
\be
\kappa  \gamma \gamma_s^3 > 3 \times 10^{13} \nu_{GHz}^3 \left( \frac{r}{R_{LC}}\right)^{3}  T_{b, 35},
\ee
which can be accommodated with the assumed multiplicity and {\Lf}s even at the light cylinder.

For (millisecond) magnetar-like parametrization (\ref{n22}), the condition (\ref{cond}) gives
\be
 \gamma \gamma_s^3 \geq \frac{ e \nu^3 r k_B T}{ m_e c^5 B \Delta \phi}= 10^7  \nu_{GHz}^3 \left(\frac{P}{10^{-3} {\rm sec}}\right)^{4} \frac{T_{b, 35}}{\Delta \phi} \left( \frac{B}{B_Q}\right)^{-1} \left( \frac{r}{R_{LC}}\right)^{4}
 \ee
Which can also be accommodated with $\gamma \sim \gamma_s \sim $ few hundred.

Finally, the ratio of the  local \Bf\ to the equipartition \Bf\ (\ref{Beq}) evaluates to
\be
b_{eq} = \frac{B_{NS} (r/R_{LC})^3}{B_{eq}}= 3 \times 10^5 \left( \frac{B_{NS}}{B_q} \right)\nu_{GHz}^{-3/2}\left( \frac{r}{R_{LC}} \right)^{-3} T_{b, 35}^{-1/2}
\label{beq}
\ee
Though Eq. (\ref{beq}) has a number of unknown parameters, there is a large region of allowed $b_{eq}  \geq 1$.

 \section{Origin of FRBs}
 
 \subsection{Not Giant Pulses from young energetic pulsars}
 \label{notGP}
 
 Having narrowed down the most likely location of the FRBs' emission to \NSs\ \mss, there are two possible  energy sources: rotation and \Bf.  \cite{2016MNRAS.462..941L} argued that if the FRBs  are analogues of giant pulses (GPs) but coming from young (ages tens to hundreds years) pulsars with Crab-like \Bf, then the required  initial periods need to be in a few msec range -  a reasonable assumption for $D \leq$ few hundreds Mpc.
 Identification of the FRB host with a galaxy at $D=1$ Gpc makes this possibility unlikely, as  we discuss next.
 
The localization of the Repeating FRB at  $\sim$ 1 Gpc \citep{2017Natur.541...58C}, an order of magnitude further away than what was a fiducial model in \cite{2016MNRAS.462..941L}, combined with a very steady value of DM (if DM was coming from the newly ejecta SN material), virtually excludes  rotationally-powered FRBs, \eg\ as analogues of Crab giant pulses \citep{2017ApJ...838L..13L}.    
   
   To reiterate the argument,  the observed radio flux  can be parametrized as a fraction $\eta\leq 1$ of the spin-down luminosity $L_{sd}$ 
\be
\nu F_\nu = {\eta} \frac{L_{sd}} {4 \pi d^2},
\ee

The longest possible spin-down time  is then
\be
 \tau_{SD} = \eta {\pi  I_{NS} \over 2 d^2  \nu F_\nu  P_{min}^2} \approx 600 \,   \eta \,  {\rm yrs}
 \label{tauSD}
 \ee
 for $F_\nu =1$ Jy and the  minimal period of $P_{min}=1 $ msec. (For a given FRB luminosity $L_{FRB}$, scaled with spin-down power, the longest  spin-down time is for shortest periods and, correspondingly, smallest \Bfs).
 The longest possible spin-down time scale (\ref{tauSD}) is barely consistent with constant value of the properties of the Repeating  FRB  over the period of few years - that would require an unrealistically high conversion efficiency $
\eta \rightarrow 1$. 

%Another constraint comes from the constant values of the dispersion measure (DM) over few years \citep{2017arXiv170101098C,2017ApJ...834L...7T}. For the Repeating FRB approximately half of the DM contribution  comes from the local plasma, $DM _{loc} \sim  100$. This is consistent with contribution from a dwarf galaxy, thus excluding large contribution from the circumburst medium \citep{2017ApJ...834L...7T}. On the other hand, as we argued above, the typical time scale of the source is very short, a SN remnant should still be present.  If DM is associated with an expanding SN shell, then it should sharply decrease with time, $DM \propto t^{-2}$   \citep{2016MNRAS.462..941L,2016ApJ...824L..32P}. Since no DM changes are seen, this further excludes young rotationally-powered pulsars as FRB sources.

Rotationally-powered radio emission - originating on the open field lines - is also disfavored due to a lack of periodicities in the Repeater(s) bursts \citep{2018arXiv181110755K}.

\subsection{Magnetar \mss?}
\label{drifts}

Repeating FRBs show similar features in their dynamic spectra: downward frequency drifts \citep{2019ApJ...876L..23H,2019Natur.566..235C,2019arXiv190803507T,2019arXiv190611305J}. \cite{2020ApJ...889..135L} interpreted these drifts as FRBs' analogues of radius-to-frequency mapping in pulsars and   Solar type-III radio burst (but not in a sense of a particular emission mechanism).  (Alternatively, spectral drifts are expected in the lensing scenarios \citep{2017ApJ...842...35C}, though in that case both upward and downward drifts are expected.)

First, we interpret similar frequency behavior as an indication of a some kind of a  stiff confining structures - most likely the magnetic field. Narrow spectral features could  then be related to the local plasma parameters (\eg\ plasma and cyclotron frequencies) or changing resonance conditions.  In both case frequency drift then reflects the propagation of the emitting region in changing magnetospheric conditions, similar to what is called ``radius-to-frequency mapping'' in pulsar research \citep[\eg][]{1977puls.book.....M,1992ApJ...385..282P}.

The frequency drift in FRBs are also reminiscent of type-III Solar radio bursts, whereby narrow frequency features show high-to-low temporal evolution \citep[\eg][]{1974SSRv...16..145F}.
  Reconnection-driven beams in magnetars \mss\ though may have different physical conditions than on the open field lines of pulsars' \mss. First, the density is not limited to the Goldreich-Julian value (\protect\ref{n11}) and can be much higher (\protect\ref{n22}). Magnetic field at the source may be higher than in pulsars (both estimates of plasma densities increase with \Bf). Finally, the bulk \Lf\ may be smaller than on the open field lines of pulsar \mss.

If the growth rate is on plasma frequency in the plasma rest frame,  $\sim \om_p'/\sqrt{\gamma}$, the condition of fast growth  in the lab frame $\om_p'/\gamma_s \geq \Omega$ translates to
\ba &
\kappa \geq { \gamma \gamma_s^3} \frac{\Omega}{\om_B} \approx 10^{-15} \gamma \gamma_s^3 \left(\frac{B_{NS}}{B_Q}\right)^{-1} \left(\frac{P}{10^{-3} {\rm sec}}\right)^{2}  \left(\frac{r}{r_{LC}}\right)^{3} &, \mbox{for scaling (\protect\ref{n11})}
\nn &
\Delta \phi \geq \gamma \gamma_s^3 \frac{\Omega^2  r}{\om_B c}  \approx 10^{-15} \gamma \gamma_s^3 \left(\frac{B_{NS}}{B_Q}\right)^{-1} \left(\frac{P}{10^{-3} {\rm sec}}\right)^{2} \left(\frac{r}{r_{LC}}\right)^{4} &, \mbox{for scaling (\protect\ref{n22})}
\label{33}
\ea
Both these constraints can generally be satisfied in \NSs\ \mss. (Same  numerical factor in front of expressions in (\ref{33}) is due to the fact that at millisecond periods
both estimates of density (\protect\ref{n11}) and (\protect\ref{n22}) coincide.)

A merger of two Langmuir waves with frequency $\sim \om_p'/\sqrt{\gamma}$ in the plasma frame will produce observed radiation at
\be
\om \sim \gamma_s \frac{\om_p'}{\sqrt{\gamma}} =
\left\{ 
\begin{array}{cc}
\left(  \kappa \Omega \om_B \frac{\gamma_s}{\gamma}\right)^{1/2} =
3 \times 10^{11} \left(  \kappa  \frac{\gamma_s}{\gamma}\right)^{1/2}  \left(\frac{r}{r_{LC}}\right)^{-3/2} {\rm rad \, s^{-1}}&, \mbox{for scaling (\protect\ref{n11})}\\
\left(  \Delta \phi  \frac{\om_B c}{r}  \frac{\gamma_s}{\gamma}\right)^{1/2} =
3 \times 10^{11}  \left( \Delta \phi  \frac{\gamma_s}{\gamma}\right)^{1/2}  \left(\frac{r}{r_{LC}}\right)^{-2} {\rm rad \, s^{-1}}&, \mbox{for scaling (\protect\ref{n22})}
\end{array}
\right.
\label{om1}
\ee
(in both cases the numerical estimates are scaled with $ \left({B_{NS}}/{B_Q}\right)^{1/2} \left({P}/({10^{-3} {\rm sec}})\right)^{-2}$.)
Both these estimates can generally produce emission at the observed radio wavelengths. 

The emission frequencies (\ref{om1}) demonstrate downward frequency drift as an emitting entity propagates up in the \NS\ \ms. For dipolar \Bf\ $\propto r^{-3}$, the scalings are $\om \propto t^{-3/2}$ and $\om \propto t^{-2}$ for the two cases (assuming constant Doppler factor; for time-varying Doppler factor $t \rightarrow t/\delta(t)^2$ ).

Let us next list a few observational  arguments for and against associating FRBs with magnetar flares. Coherent radio emission can be produced at the initial stage of a ``reconnection flare'', whereby coherent ``kinetic jets'' of particles are generated, like the ones in the studies of Crab flares \citep[\eg][]{2014PhPl...21e6501C,2017JPlPh..83f6301L,2018JPlPh..84b6301L}. But there are observational constraints: (i)  the SGR 1806 - 20 flare had peak power of $ 10^{47}$ erg s$^{-1}$  \citep{palmer} but was not seen by Parkes radio telescope \citep{2016ApJ...827...59T}; that puts an upper limit on radio-to-high -energy efficiency $\leq 10^{-6}$. For the Repeater, the first Repeater,  the implied high energy luminosity would be then $\geq 10^{47}$ erg s$^{-1}$. On the other hand,
if the  Repeater was in our Galaxy the corresponding fluxes would be in GigaJansky, which are clearly not seen. 
Also in case of PSR J1119-6127 magnetar-like X-ray bursts seem to suppress radio emission \citep{2017ApJ...849L..20A}, but this is probably related to rotationally-driven radio emission, not reconnection-driven. Thus, only some special types of magnetars  can produce FRBs.

Finally, associating FRBs with (special kinds of) magnetar flares may resolve the lack  of periodicities in the appearance of the FRBs from the Repeater(s): magnetospheric reconnection events appear randomly on closed field lines. This, combined with short, millisecond-like periods, will likely erase the signatures of the rotational period in the observed time sequence of the bursts.

\section{Discussion}

In this paper we discuss the constraints on the properties of coherent emitting astrophysical  sources, having in mind relativistic objects like pulsars and, presumably, FRBs. Several lines of reasoning  point to {\NS}s' \mss\ as the origin of the high brightness emission (this is of course known for pulsars, but is  important  for FRBs). 

An important point, besides the estimates of plasma parameters at the source for given observed brightness, is the estimate of non-coherent energy losses,  \S
\ref{a1}. 
Let us discuss this important argument. Particles at the source lose energy to emission of coherent (resonant)  waves and, in addition, to non-resonant interactions (\eg\ IC and synchrotron). Typically, the part of the energy that goes to  coherent  (low frequency - radio) emission is much smaller than the one going to non-resonant interactions (often in optical and X-rays). This was not much of a problem before the identification of FRBs at cosmological distances - the radio emission was always energetically unimportant,  subdominant part. For example, even the brightest and rarest giant pulses from Crab reach only $10^{-2}$ of the total spin-down  luminosity. 
Observations of FRBs, raise the bar, so to say. The implied radio luminosity is some five to nine {\it orders of magnitude larger} that is seen in pulsars (In Crab the average radio power is $\sim 10^{32}$ erg s$^{-1}$, the peak   is $\sim 10^{36}$ erg s$^{-1}$; the FRB Repeater is at  $\sim 10^{41}$ erg s$^{-1}$). These are macroscopically (in astrophysical sense) important powers \citep{2017ApJ...838L..13L} and  thus do offer, for the first time from radio observations, a meaningful physical constraints on the plasma  parameters at the source. We demonstrated that though these constraints are important, they can be realistically satisfied. 

In conclusion, FRB emission properties point to \mss\ of \NSs\ as the origin. Two types of mechanisms can be at work - rotationally or magnetically powered. Rotationally-powered FRB emission mechanisms \citep[\eg as analogues of Crab giant pulses][]{2016MNRAS.462..941L} are excluded by the  localization of the Repeating FRB at $\sim 1$ Gpc  \citep{2016Natur.531..202S}, as discussed by \cite{2017ApJ...838L..13L} and \S \ref{notGP}. Magnetically-powered emission has some observational constraints, as discussed at the end of \S \ref{drifts}, but remains theoretically viable.

  %%%%%%%%%%%%%%%%%%%%%%%%%%%%%%%%%%%%%%%%%%%%%%%%%%%%%%%%%%%%%%%%%
\section*{Acknowledgments}
This work had been supported by DoE grant DE-SC0016369 and
NASA grant 80NSSC17K0757. We would like to thank organizers of  SRitp  FRB workshop for hospitality, and particularly discussions with Jim Cordes, David Eichler, Victoria Kaspi, Jonathan Katz, Amir Levinson, Yuri Lyubarski and Ue-Li  Pen. We would like to thank Maxim Barkov  for discussions.

%%%%%%%%%%%%%%%%%%%%%%%%%%%%%%%%%%%%%%%%%%%%%%%%%%%%%%%%%%%%%%%%%

%\bibliographystyle{apj}
%\bibliographystyle{apsrev}

\bibliographystyle{jpp}

  \bibliography{/Users/maxim/Home/Research/BibTex}

\appendix

\section{Brightness Temperature and radiation energy-density}
\label{Tb}
\subsection{Isotropic source}
\label{app:sec:isotropic}

The energy $ dE $ from within a solid angle $ d\Omega $ passing through a projected area $ dA $ in time $ dt $ and in a narrow frequency range $ d\nu $ is
\begin{equation}\label{eq:dE}
    dE = I_\nu\, dA\, dt\, d\Omega\, d\nu,
\end{equation}
where $ I_\nu $ is the specific intensity (or spectral brightness). The intensity  $ I_\nu $ is a  conserved quantity along the ray path.

The spectral flux density $ F_\nu $, power emitted per unit area per unit frequency, is given by
\begin{equation}\label{eq:Fnu}
    F_\nu = \int I_\nu \cos\theta\, d\Omega,
    \quad {\rm with}\quad
    d\Omega = \sin\theta\, d\theta\, d\phi,
\end{equation}
using the effective area $ \cos\theta\, dA $ in~\eqref{eq:dE}.

The flux density $ S_\nu $, the spectral power received per unit area per unit frequency by a detector from a discrete source (i.e. one for which there is a well-defined solid angle) subtending a solid angle $ \Omega_{\rm source} $ is
\begin{equation}\label{eq:Snu}
    S_\nu = \int_{\Omega_{\rm source}} I_\nu \cos\theta\, d\Omega,
\end{equation}
%with $ [S_\nu] = {\rm J}\cdot{\rm s}^{-1}\cdot{\rm m}^{-2}\cdot{\rm Hz}^{-1} $. 
For an isotropic source of radius $ R $ (i.e. a sphere of uniform spectral brightness) located at a distance $ d $ from the observer we obtain
\begin{equation}\label{eq:Snu2}
    S_\nu = I_\nu \int_0^{2\pi} d\phi \int_0^{\theta_{\rm c}} d\theta\, \cos\theta\sin\theta = \pi I_\nu \sin^2\theta_{\rm c} = \pi I_\nu \frac{R^2}{d^2},
\end{equation}
where $ \theta_{\rm c} = \sin^{-1}(R/d) $. In particular $ S_\nu = \pi I_\nu $ for $ d  = R $.

 We comment that $ I_\nu $ cannot be obtained for unresolved sources since $ \theta_{\rm c} $ cannot be measured.
{If we know the distance and variability time, then we can estimate  $ \theta_{\rm c} $}.

Planck's law states that for blackbody radiation $ I_\nu = B_\nu $ with
\begin{equation}\label{eq:BB}
    B_\nu = \frac{2h\nu^3/c^2}{\exp(h\nu/k_BT) -1} \approx \frac{2 k_B T \nu^2}{c^2},
\end{equation}
where the approximation applies when $ h\nu/k_B T \ll 1 $. With a substitution $T \rightarrow T_b$ Eq. (\ref{eq:BB}) defines the brightness temperature. 

 For an opaque (can receive radiation from only one hemisphere) isotropic source with spectral brightness $ I_\nu $, the spectral flux density $ F_\nu $ is
\begin{equation}\label{eq:Fnu1}
    F_\nu = I_\nu\int_0^{2\pi}d\phi\int_0^{\pi/2}d\theta\, \cos\theta\sin\theta = \pi I_\nu.
\end{equation}
Comparing~\eqref{eq:Snu2} and~\eqref{eq:Fnu1} we note that for an isotropic source $ F_\nu = S_\nu $ on the surface of the source; they are not equal otherwise. 

For isotropic blackbody radiation this implies that
\begin{equation}\label{eq:FnuSnu}
    F_\nu = \pi I_\nu = \pi B_\nu,\quad
    S_\nu = \pi I_\nu \frac{R^2}{d^2} = \pi B_\nu \frac{R^2}{d^2},
    \quad{\rm and}\quad
    F_\nu = S_\nu \frac{d^2}{R^2}.
\end{equation}

We may write $ dE $ in~\eqref{eq:dE} as
\begin{equation}\label{eq:dE2}
    dE = c u_\nu(\Omega)\, dA\, dt\, d\Omega\, d\nu,
\end{equation}
where $ u_\nu(\Omega) $ is the spectral energy density per unit solid angle per unit frequency.
% with $ [u_\nu(\Omega)] = {\rm J}\cdot{\rm m}^{-3}\cdot{\rm sr}^{-1}\cdot{\rm Hz}^{-1} $. 
Using~\eqref{eq:dE} and~\eqref{eq:dE2} we obtain the relationship
\begin{equation}
    I_\nu = c u_\nu (\Omega).
\end{equation}
The spectral energy density $ u_\nu $, energy per unit volume per unit frequency, is obtained by integrating over solid angle
\begin{equation}\label{eq:unu}
    u_\nu = \int u_\nu(\Omega)\, d\Omega = \frac{1}{c}\int I_\nu\, d\Omega,
    \quad\text{implying}\quad
    u_\nu = \frac{4\pi}{c}I_\nu
\end{equation}
for an isotropic source.
% with $ [u_\nu] = {\rm J}\cdot{\rm m}^{-3}\cdot{\rm Hz}^{-1} $.

The brightness temperature of an isotropic source may be defined using~\eqref{eq:BB}, \eqref{eq:FnuSnu} and~\eqref{eq:unu} as
\begin{equation}\label{eq:kBT1}
    k_B T_b = \frac{c^2 I_\nu}{2 \nu^2} 
        = \frac{c^2 F_\nu}{2\pi \nu^2}
        = \frac{c^2 S_\nu}{2\pi \nu^2}\frac{r^2}{R^2}
        = \frac{c^3 u_\nu}{8\pi \nu^2 }.
\end{equation}
We may integrate $ I_\nu $, $ F_\nu $, $ S_\nu $ and $ u_\nu $ over the observation frequency range $ \nu_1 \leq \nu \leq \nu_2 $ to obtain, respectively,
\begin{equation}
    I = \frac{2 k_B T_b \nu^3}{c^2}\frac{\Delta\nu}{\nu},\quad
    F = \pi I,\quad
    S = \pi\frac{R^2}{r^2} I,\quad
    u_r = \frac{4}{c}I,
\end{equation}
where we define $ \nu = (\nu_1 + \nu_2)/2 $ and $ \Delta\nu = \nu_2 - \nu_1 $, and assume that $ \Delta\nu/\nu \ll 1 $ so that $ \nu_2^3 - \nu_1^3 \approx 3\nu^3(\Delta\nu/\nu) $. The brightness temperature may then be expressed as
\begin{equation}\label{eq:kBT5}
    k_B T _b= \frac{c^2I}{2\nu^3}\frac{\nu}{\Delta\nu} 
        = \frac{c^2 F}{2\pi \nu^3}\frac{\nu}{\Delta\nu} 
        = \frac{c^2 S}{2\pi \nu^3}\frac{\nu}{\Delta\nu}\frac{r^2}{R^2}
        = \frac{c^3 u_r}{8\pi \nu^3}\frac{\nu}{\Delta\nu}.
\end{equation}

\subsection{Unresolved source}
\label{app:sec:unresolved}

The emitted spectral density, $ F_\nu $, and received spectral density, $ S_\nu $, are defined in terms of specific intensity, $ I_\nu $, but with the integral over solid angle performed over different domain sizes. For a resolved isotropic sources relations (\ref{eq:FnuSnu}) apply. 
For an unresolved source $ F_\nu $ and $ S_\nu $ cannot be related through $ I_\nu $ as $ \theta_{\rm c} $ cannot defined. However, if we are able to infer the size of the source through another mean then it is possible to relate $ S_\nu $ and $ F_\nu $. Suppose an isolated (i.e. radiation arriving at the detector is only from the source), unresolved, isotropic source has an inferred radius of $ R_{\rm th} $  (\eg\ through variability time scale $ R_{\rm th} \sim \tau c$) and is at a distance $ d $ from the observer. Then its angular size is given by $ \theta_{\rm c, th} = \sin^{-1}(R_{\rm th}/r) $. For an unresolved source we make the replacement $ \theta_{\rm c} \to \theta_{\rm c, th} $ in~\eqref{eq:kBT1} and~\eqref{eq:kBT5}. Of course we have $ \theta_{\rm c} = \theta_{\rm c, th} $ for resolved sources.

\subsection{Anisotropic source}
\label{App:sec:anisotropic}

Pulsars' (and possibly FRBs) emission is beamed and hence anisotropic at the source. Let us assume that within the beam, there are no anisotropies in frequency.
%\footnote{MZR: I find it difficult to justify any model for spectral anisotropy of an unresolved source region.} 
We may approximate the anisotropy due to beaming by restricting the emission region to $ 0\leq \theta \leq \theta_{\rm b} $: emission is from a circular patch of the source surface with angular extent $ \theta_{\rm b} $. For blackbody radiation we may write this as $ I_\nu(\theta) = 2k_BT(\theta) \nu^2/c^2 $ with
\begin{equation}\label{eq:aniso}
    T_b(\theta) = 
    \begin{dcases}
    T_b, & \quad {\rm for}\quad \theta \leq \theta_{\rm b},\\
    0, & \quad {\rm otherwise},
    \end{dcases}
\end{equation}
which implies that
\begin{equation}
    F_\nu = \frac{2k_BT\nu^2}{c^2}\int_0^{2\pi}d\phi\int_0^{\theta_{\rm b}}d\theta\, \cos\theta\sin\theta
        % = \frac{2\pi\sin^2\theta_{\rm b}k_BT\nu^2}{c^2}
        % = \pi\sin^2\theta_{\rm b}I_\nu
        \approx \pi\theta_{\rm b}^2I_\nu,
\end{equation}
where the final expression is valid for $ \theta_{\rm b} \ll 1 $, and $ I_\nu = B_\nu $ is given by~\eqref{eq:BB}. Similarly,
\begin{equation}
    u_\nu 
        % = \frac{1}{c}I_\nu \int_0^{2\pi}d\phi\int_0^{\theta_{\rm b}}d\theta\, \sin\theta
        = \frac{2\pi}{c}(1-\cos\theta_{\rm b})I_\nu
        \approx \frac{\pi\theta_{\rm b}^2}{c}I_\nu,
    \quad{\rm and}\quad
    S_\nu = \pi\theta_{\rm b}^2\frac{R_{\rm th}^2}{r^2}I_\nu,
\end{equation}
we approximate $ \theta_{\rm c, th} $ using $ \tan\theta_{\rm c, th} = R_{\rm th} \sin\theta_{\rm b}/r $ to obtain $ \theta_{\rm c, th} \approx (R_{\rm th}/r)\theta_{\rm b} $, where we assume $ \{\theta_{\rm c, th}, \theta_{\rm b}\} \ll 1 $. Therefore, for an anisotropic source described by~\eqref{eq:aniso} the brightness temperature as given by~\eqref{eq:kBT1} and~\eqref{eq:kBT5} is modified to
\begin{equation}
    k_BT _b
    = \frac{c^2 I_\nu}{2\nu^2}
    = \frac{c^2 F_\nu}{2\pi\theta_{\rm b}^2\nu^2}
    = \frac{c^2 S_\nu}{2\pi\theta_{\rm b}^2\nu^2}\frac{r^2}{R_{\rm th}^2}
    = \frac{c^3 u_\nu}{2\pi\theta_{\rm b}^2\nu^2},
\end{equation}
and
\begin{equation}
    k_BT _b 
    = \frac{c^2 I}{2\nu^3}\frac{\nu}{\Delta\nu}
    = \frac{c^2 F}{2\pi\theta_{\rm b}^2\nu^3}\frac{\nu}{\Delta\nu}
    = \frac{c^2 S}{2\pi\theta_{\rm b}^2\nu^3}\frac{\nu}{\Delta\nu}\frac{r^2}{R_{\rm th}^2}
    = \frac{c^3 u_r}{2\pi\theta_{\rm b}^2\nu^3}\frac{\nu}{\Delta\nu}.
    \label{eq:final}
\end{equation}

\section{Coordinate and observer times, and mini-jets}
\label{times}

 Let us clarify here the often confusing notions of coordinate times  in two systems  of references and the laboratory  observer time. Relation $\tau =\delta ^{-1} \tau' $ is between the {\it observer} time $\tau$ in our, the laboratory, frame and the coordinate time $\tau' $ in the plasma rest frame (which is the same as the observer time in the plasma frame). The coordinate time in the plasma rest frame is related  to the coordinate time in  the laboratory frame $\tau_{coord}$ as $\tau' = \tau_{coord}/\gamma_s$. Thus,  $\tau = \tau_{coord}/ (\delta \gamma_s) \approx  \tau_{coord}/ (2 \gamma_s^2)$, where the last relation approximates $\delta  \approx 2  \gamma_s$ for relativistic motion along the line of sight. Thus, the relations (\ref{LF})
  do include the commonly used notion that  a relativistically   moving  emitter that is active for (coordinate time) $ \tau_{coord}$   produces  a pulse in observer time which is $2 \gamma_s^2$ shorter. 

Along the similar lines of thought,  the coherently emitting entity can be a type of  mini-jet propagating within the emission regions with internal \Lf\ $\gamma_j$. The causally connected region, that can in principle supply energy to the jet with   duration  $\tau'$ in the frame of the blob 
 is then $\sim \pi   (c \tau')^3 \gamma_j^2 $ \citep{2006MNRAS.369L...5L}. In the observer frame the pulse will still have a duration $\tau =\delta ^{-1} \tau' $.

Hence relation (\ref{05}) will be modified as
\be
{k_B T_b} = 
\eta_r {\gamma \lambda ^3  m_e c^2 n' } \delta^{4} \gamma_j^2
\label{06}
\ee
Thus, minijets will further reduce demands on the intrinsic energy density at the source, by $\sim \gamma_j^2$.

\end{document}